\documentclass[sigconf, nonacm]{acmart}

\AtBeginDocument{%
  }

\setcopyright{none}
\renewcommand\footnotetextcopyrightpermission[1]{} 
\usepackage{subfig}
\usepackage{stfloats}
\usepackage{color,soul}
\usepackage{multirow}
\usepackage{longtable}
\usepackage{appendix}
\begin{document}

\title{Evaluating Synthetic Command Attacks on Smart Voice Assistants}


\author{Zhengxian He}
\affiliation{%
  \institution{Georgia Institute of Technology}
  \city{Atlanta}
  \country{USA}}
\email{zhe72@gatech.edu}

\author{Ashish Kundu}
\affiliation{%
  \institution{Cisco Research}
  \city{San Jose}
  \country{USA}
}
\email{ashkundu@cisco.com}

\author{Mustaque Ahamad}
\affiliation{%
\institution{Georgia Institute of Technology}
  \city{Atlanta}
  \country{USA}}
\email{mustaq@cc.gatech.edu}


\begin{abstract}
Recent advances in voice synthesis, coupled with the ease with which speech can be harvested for millions of people, introduce new threats to applications that are enabled by devices such as voice assistants (e.g., Amazon Alexa, Google Home etc.). We explore if unrelated and limited amount of speech from a target can be used to synthesize commands for a voice assistant like Amazon Alexa. More specifically, we investigate attacks on voice assistants with synthetic commands when they match command sources to authorized users, and applications (e.g., Alexa Skills) process commands only when their source is an authorized user with a chosen confidence level. We demonstrate that even simple concatenative speech synthesis can be used by an attacker to command voice assistants to perform sensitive operations. We also show that such attacks, when launched by exploiting compromised devices in the vicinity of voice assistants, can have relatively small host and network footprint. Our results demonstrate the need for better defenses against synthetic malicious commands that could target voice assistants.

\end{abstract}

\ccsdesc[500]{Security and privacy~Embedded systems security}
\ccsdesc[300]{Security and privacy~Biometrics}
\ccsdesc[300]{Hardware~Sound-based input / output}

\keywords{Voice Assistants, Speech Analysis and Synthesis, Speaker Verification, Access Control}


\maketitle
\section{Introduction}
Voice is a natural way for people to interact with devices in their vicinity. It is one of the reasons for the increasing adoption of voice assistants such as Amazon Alexa, Google Assistant and Samsung Bixby. At the same time, applications enabled by such voice assistants are increasing at a rapid pace as can be seen by the diversity and number of Amazon Skills. The proliferation of voice assistants and applications supported by them leads to new security threats. In fact, researchers have explored how voice assistants can be targeted by malicious commands that can be issued when the attacker is in close physical proximity or across the network~\cite{diao2014your, jang2014a11y, Carlini:2016vl, schonherr2018adversarial, Carlini:2018wj, Abdullah:2018ho, Yuan:2018um, kumar2018skill, qin2019imperceptible}. Others have explored thousands of applications supported by voice assistants and the sensitive actions performed by them~\cite{shezan2020read}.

We explore if attackers can leverage the ease with which speech can be harvested to launch attacks against voice assistant enabled applications. 
As with email addresses and phone numbers, limited and unrelated speech can be easily harvested for a large number of people. By unrelated speech, we mean speech that does not include words that must be uttered for specific voice assistant commands. For many users, unrelated speech may be available from podcasts, YouTube videos, lectures, talks, online posts, or can even be collected by making robocalls. 


A possible defense against malicious commands is to use the command voice itself to determine if the command is coming from an authorized user. To enable this, authorized users must set up their profiles to allow command voice to be matched against the voices of users having profiles.  
However, because of a variety of reasons, including usability and environmental constraints (e.g., noisy background, distance between speaker and voice assistant), a command source's match with an authorized user voice is not required to be strict. In the context of malicious voice commands, to the best of our knowledge, the efficacy of such defenses in current voice assistants has not been explored.

Although widely available services can now be used for speech synthesis~\cite{ElevenLabs2023, Coqui2023, wang2023neural}, attackers will prefer to avoid them for cost or detection reasons when targeting a large number of victims. Since attack commands that target voice assistants may not require natural-sounding human quality speech, we explore the feasibility of low cost techniques for speech synthesis which allow such synthesis to be done even at compromised user devices. 
We have two specific goals for how commands are synthesized. First, the synthesized command needs to be {\it intelligible} to the voice assistant so it is recognized by it. Second, the command must preserve {\it similarity} with an authorized user's voice so it can work even when the voice assistant enabled application matches commands to users with a certain level of confidence. If these two goals can be met for a significant fraction of targeted users and their voice assistants, it will demonstrate the feasibility of low cost and large scale attacks on voice-enabled applications.

We present an empirical security analysis for commands of various lengths when applications check that the command voice comes from an authorized user. To conduct experiments at scale to compare both intelligibility and similarity of synthetically generated commands, we develop an experimental testbed with a popular voice assistant (e.g., Amazon Alexa). We use an efficient and lightweight concatenative speech synthesis scheme to generate attack commands. In particular, we use a unit-selection method that extracts diphones from available speech and concatenates necessary diphones to synthesize a command~\cite{lenzo2000diphone, OSHAUGHNESSY198855}. Our automated testbed allowed us to conduct over one thousand experiments for many commands and user profiles. The results of these experiments help us demonstrate the following:

\begin{enumerate}

\item We show that when available target speech has all diphones needed for the synthesis of a command, the Alexa voice assistant correctly recognizes 93.8\% of the commands generated with a basic unit-selection concatenative synthesis method. Our experiments used both short and long commands in the voice of a diverse group of users (e.g., accents, gender etc.).
\item Similar to command recognition, we found that unit-selection concatenative synthesis also preserves voice similarity as assessed by the speaker match confidence level. In our experiments, the highest confidence level in speaker voice similarity was returned for 90\% of the users who have profiles that vary in accent and gender. 
Thus, our results show security vulnerabilities of voice assistants to synthetic commands even when applications match command source with authorized users.
 \item 
 We demonstrate that 50\% of the commands can successfully activate a voice assistant when synthesized from only 30 seconds of unrelated speech of a target. This is true even when applications processing these commands require a high level of confidence in the similarity of the command source voice with the voice of an authorized user. We show that the success rate increases as more speech becomes available, reaching 80\% with 4 minutes of target speech.    
\end{enumerate}

Our results demonstrate the ease with which voice assistant enabled  applications can be targeted by harvesting speech and efficiently synthesizing attack commands. To the best of our knowledge, we are the first to show that voice profile matching, as used currently, provides little protection against such malicious commands. 

The paper is structured as follows. Section~\ref{related} discusses related work and the threat model is presented in Section~\ref{threat}. We discuss our approach in Section~\ref{approach} and the system developed for carrying out experiments is described in Section~\ref{system}. The results of our experiments are discussed in detail in Section~\ref{evaluation}. The paper is concluded with discussions and conclusions in Sections~\ref{discussion} and ~\ref{conclusions}.

\section{Related Work}
\label{related}
Voice assistant attacks have been explored extensively in past research which has been surveyed in~\cite{10.1145/3527153}. Early research explored hidden malicious commands that target voice assistants which can be processed by them but not heard by people who are close to them. Since voice assistants rely on speech recognition to infer commands, several important works have explored how commands can be made inaudible to people while ensuring that they can still be correctly recognized as intended commands~\cite{diao2014your, jang2014a11y, Carlini:2016vl, schonherr2018adversarial, Carlini:2018wj, Abdullah:2018ho, Yuan:2018um, kumar2018skill, qin2019imperceptible}. Notably, researchers such as Yuan et al.~\cite{Yuan:2018um} and Carlini et al.~\cite{Carlini:2016vl,Carlini:2018wj} have demonstrated the feasibility of these attacks using adversarial examples. Other methods like ultrasonic sound~\cite{Abdullah:2018ho,Roy:2018ws,Zhang:2017kl, Yan:2020fa} have also been explored to make commands inaudible. Researchers have also demonstrated attacks that inject malicious commands directly into microphones on smartphones using electromagnetic interference (EMI) generated by wireless chargers~\cite{dai2022inducing}. 
Others have explored how
voice assistant vulnerability exploitation can lead to consequences such as exfiltration of sensitive data from compromised computers~\cite{he2021compromised}. 
 However, it's worth noting that these studies did not  address synthesized malicious commands  when voice assistants check voice similarity with authorized users.

There is considerable body of speech synthesis research which has explored numerous techniques, including recent use of deep neural networks for high-quality speech synthesis (see ~\cite{tan2021survey} for a survey of speech synthesis). These techniques vary in the amount of speech needed for a target, the quality of the synthesized speech and other resources required by speech synthesis. Since we synthesize commands that target voice assistants, our goal is not to produce high-quality speech that humans find indistinguishable from real speech. Also, to target a large number of users, it is desirable that commands can be generated from speech without specific text constraints like appearance of certain words. 

Although high-quality speech synthesis, which converts text to speech, can be used to synthesize malicious commands that target voice assistants, attacks using such speech synthesis must train text-to-speech models~\cite{kim2021conditional} or must utilize application programming interfaces (APIs) of services implemented in the Cloud. A substantial amount of speech data may be required by high-quality speech synthesis models and they can be resource-intensive. Some researchers have proposed novel methods for low-resource text-to-speech synthesis, achieving good results even with limited speech data~\cite{huybrechts2021low, xu2020lrspeech, gabrys2022voice}. For instance, Xu et al.~\cite{xu2020lrspeech} achieved good results with only one minute of original speech from the target. However, these speech resource-efficient methods have high memory and processing footprints.

We explore attacks even when voice assistants check similarity of command speech against authorized user profiles and verify call sources. Speaker verification attacks can be carried out by using voice cloning, voice conversion or speech synthesis~\cite{li2023freevc, kim2021conditional,huybrechts2021low, xu2020lrspeech, gabrys2022voice}. Generative AI enabled voice cloning services are now available but they strive to generate voice for human users. As a result, they either require more speech for a target or must utilize models with high memory and network footprint.
Voice conversion attacks, which create a speaker model for a victim from speech, can convert speech from another user to sound like it is from the victim. However, such attacks require both the source and target speaker models and could be resource-intensive for large-scale attacks. 

Advances in voice synthesis have made it possible to synthesize realistic-sounding speech for a targeted user with a limited amount of speech. Liveness detection solutions that check if a command is synthetic and not coming from a human user have been proposed~\cite{blue2018hello} but they are currently not employed by voice assistants. Furthermore, recent research has reported successful attacks against methods that rely on detection of synthetic speech~\cite{kassis2023breaking}. 

In this study, we explore the feasibility of launching attacks against voice assistants by utilizing a limited amount of unrelated speech data as well as computing power and storage resources that are commonly available on a victim's compromised computer. Furthermore, we do not want a compromised device to download large models, as this may trigger network defenses. Similarly, attack code should avoid accessing remote service APIs that can raise suspicion. Our aim is to assess whether low cost speech synthesis techniques that do not have such requirements can be used to get voice assistants to execute commands even when they match the command source voice to authorized users. Based on these requirements, we chose concatenative speech synthesis which utilizes a unit-selection method to generate commands from unrelated and limited speech.
We discuss the details of this unit-selection synthesis method and its efficacy in the remainder of this paper.

\section{Threat Model}
\label{threat}
In the context of our threat model, an attacker's aim is to remotely launch attacks on a large number of voice assistants belonging to many users. This attack involves injecting a series of synthetically generated audio commands through speakers of compromised devices which are in the vicinity of targeted voice assistants. 
%
As a concrete example, we envision a scenario where an attacker successfully compromises a victim's personal computer (PC) and installs malware on it. 
The attacker can leverage common cybersecurity vulnerabilities such as software flaws or even use social engineering tactics~\cite{malware}. As an additional requirement of this threat model, we assume that the infected machine is located within the vicinity of a voice assistant, a situation that is increasingly common considering the ubiquity of voice assistant devices in our homes and offices~\cite{Vigliarolo:2017,Libert:2017}.

The envisioned attack can be launched at scale when the attacker is able to target many users. Ideally, the attack must blend in with normal activity, thus reducing the chances of detection and mitigation via common network defenses. 

\subsection{Collecting Target User Speech} 
We consider attacks where voice assistants match command source voice to an authorized user. Speech from an authorized user is needed to synthesize commands that can meet this requirement. Recorded speech of a targeted user can be collected from numerous sources. For example, speech could be harvested for many users from public sources like Youtube videos, podcasts, even lectures or talks. In the absence of such sources, a user's phone can be targeted by robocalls and the conversation of those who respond to the call can be recorded. Finally, a targeted user's voice can be recorded stealthily with the built-in microphone of a compromised computer which we assume is in the vicinity of the voice assistant. This last option could be more attractive because harvesting of speech samples needed for the attack can be done with minimal network activity. In all cases, it is desirable that only a \textbf{limited} amount and \textbf{unrelated} audio recordings of victims are used. However, these recordings should enable the malware to synthesize malicious commands in the targets' voices with a high likelihood of a successful attack.

\subsection{Low Network and Host Activity} We assume that the malware on the compromised system avoids excessive network usage that can be easily detected. This assumption is based on the observation that in most modern environments - be it business or home settings - some form of \textit{Intrusion Detection and Prevention System (IDPS)} is in use~\cite{scarfone2012guide}. Such a system can detect and block any suspicious communication when the network activity appears to be highly anomalous.
Similarly, a substantial drain on a host's computational resources or noticeable increase in data usage can raise alerts. To remain undetected, the malware should be designed to be resource-efficient, operating under the radar of common resource-monitoring systems.

\subsection{Black Box Analysis} We want to explore attacks that do not require access to the internal workings or intermediate outputs of voice assistant models, which is possible in white box attacks~\cite{li2020practical, xie2020real, marras2019adversarial, nakamura2019v2s,saito2017training, saito2017statistical}. Also, unlike~\cite{tian20_odyssey} in black box settings, we want to exclude making numerous queries to the model, as this could significantly increase the chance of detection. Our goal is to construct an attack that is effective irrespective of the underlying models used by the voice assistant to process commands.
Due to these constraints, highly sophisticated text-to-speech models - which often require substantial data and computational resources - are not considered by us. Instead, we utilize a lightweight and smaller resource footprint approach for synthesizing commands. 


\section{Approach}
\label{approach}
We follow an empirical security analysis approach that targets real voice assistants with synthetic commands. Our approach includes (i) choosing a target victim and creating the victim's profile on the voice assistant, (ii) acquiring limited and unrelated speech for the target, (iii) synthesizing attack commands from unrelated speech, and (iv) testing if the synthetic commands are successfully processed by the voice assistant that is set up with the victim profile. This is repeated for a sufficient number of users and commands to assess the efficacy of such an attack. To achieve this goal, we first discuss our approach for designing the experiments and how we run them on the testbed discussed in Section~\ref{system}.

\subsection{Target Speech Collection}
In an actual attack scenario, a victim's profile will already exist on the targeted voice assistant. However, to conduct experiments that target victim {\it A}, we first need to set up a profile for {\it A} on the voice assistant. To avoid privacy concerns related to voice-biometric data, we chose a well-trained and high quality text-to-speech (TTS) system to generate commands to set up a victim profile in our experiments. As for the unrelated speech which is used to synthesize attack commands, we use actual speech.

We utilized a pre-trained text-to-speech model provided by Coqui TTS~\cite{Coqui2023}, specifically, VITS~\cite{kim2021conditional}, trained on the VCTK Corpus~\cite{veaux2017cstr}. Coqui TTS provides implementations for various TTS models, from spectrogram models to end-to-end models. It also provides tool sets for users to fine tune or train their own models. VITS is an end-to-end TTS model that can achieve an audio quality Mean Opinion Score (MOS) of 4.49, comparable to ground truth, ranked 2nd in a TTS synthesis LJSpeech benchmark~\cite{paperswithcode_ljspeech_2023}. VCTK Corpus is a comprehensive dataset which encapsulates speech data from a diverse set of 110 English speakers, each contributing their unique accent. Each speaker has rendered approximately 400 sentences derived from various sources. This corpus is particularly suited for text-to-speech synthesis systems, especially for speaker-adaptive systems and neural waveform modeling \cite{yamagishi2019cstr}. For the purposes of our study, we chose 10 male voices and 10 female voices. Our chosen voices include different accents from different English-speaking regions. Each of these voices defines a user profile in our experiments.

Coqui TTS allowed us to generate commands for setting up profiles in our experiments. Its training dataset, VCTK Corpus, provided us the unrelated actual speech for chosen voice profiles that are the attack targets in the experiments. This unrelated speech serves as input to our low cost speech synthesis method that generates attack commands. 

\subsection{Extraction of Words and Diphones}
\begin{figure}[ht]
    \centering
    \includegraphics[width = 0.4\textwidth]{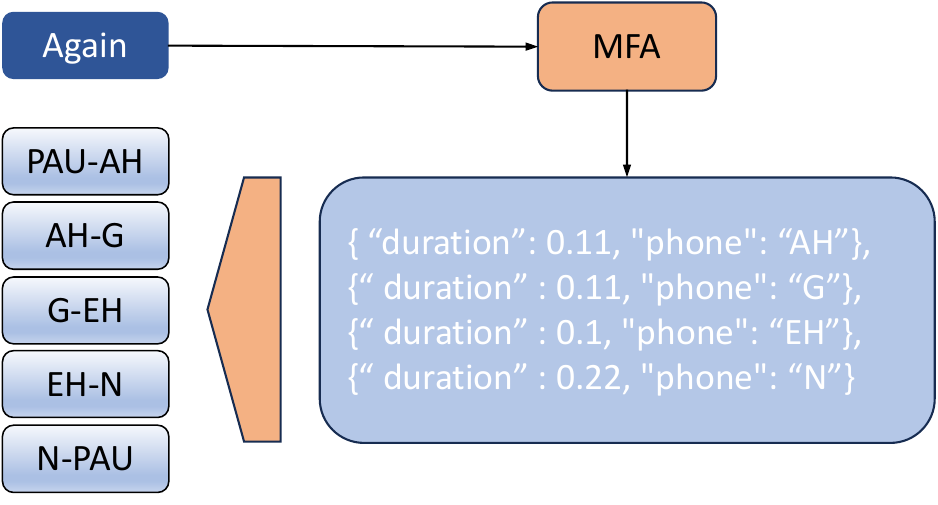}
    \caption{Diphone Extraction of Word ``Again'' Using MFA.}
    \label{fig:mfa}
\end{figure}

A phoneme is the basic sound unit in speech and forms the building blocks of word sounds. For example, the word ``Again'' consists of 4 phonemes, ``AH'',``G'',``EH'' and ``N'' as in ARPAbet phonetic transcription codes~\cite{enwiki:1103994253}.
A diphone is a linguistic unit that is derived from a pair of adjacent phonemes in speech. In the word ``Again'', 4 phonemes can be used to form 3 diphones. In our work, silence is treated as a phoneme as it is integral to creating pauses between words, denoted as ``PAU''. Consequently, we will incorporate two additional diphones at the beginning and end. This concept is visually illustrated in Fig.~\ref{fig:mfa}. 


To extract words and diphones from any given set of audio files, we leverage the forced aligner, \textit{Montreal Forced Aligner (MFA)}. Based on the Hidden Markov Model and built on Kaldi, MFA is able to provide precise timing information for each word and phoneme in a given media file \cite{mcauliffe17_interspeech}. Leveraging the timing for each word and phoneme, we can effectively cut the word and phonemes from the media file and store them for command synthesis. Fig.~\ref{fig:mfa} provides an example output of MFA.

Our unit-selection based speech synthesis method uses diphones to synthesize commands. As suggested in past work, we use diphones that are obtained by dividing a speech waveform into phone-sized units, with cuts in the middle of each phoneme~\cite{OSHAUGHNESSY198855, darbandi2002speech}. Thus, if $p_1$ and $p_2$ are two adjacent phonemes, a diphone is extracted from them by combining second half of $p_1$ with the first half of $p_2$. We intentionally omit the initial half of the first phoneme and the second half of the final phoneme. 
For words, we follow a similar approach, leaving space at the start and end for half a phoneme, ensuring they can be easily combined with other linguistic units. 

\subsection{Attack Command Synthesis}
\label{section:command_synthesis}
Once we have words and diphones from the speech of a targeted victim, we then generate the desired attack command. Initially, we search for words that appear in the command within our set of extracted words from the target's unrelated speech. This usually results in a handful of commonly occurring words that can be directly incorporated into the output. Subsequently, we generate the remaining words in the command by concatenating their diphones in order. Based on the missing words in a command, we can compile a list of diphones needed to synthesize these words. This list is then cross-referenced with our extracted diphones. If all the necessary diphones for a word in the command are present, we concatenate the found diphones and words. If any diphones are missing, we substitute them with words or diphones from a voice profile that is likely to bear a resemblance to the target's voice (in our experiments, we choose diphones from a user profile with the same gender). 

\begin{figure}[htbp]
    \centering
    \includegraphics[width = 0.3\textwidth]{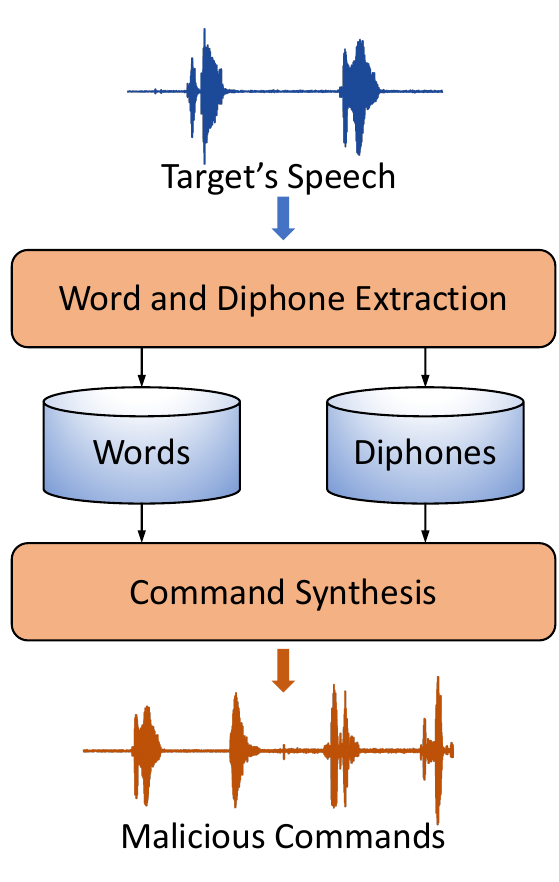}
    \caption{Unit-Selection Synthesis.}
    \label{fig:synthesis}
\end{figure}

\section{System Architecture and Implementation}
\label{system}
The task of evaluating whether a command generated with our approach is processed by a voice assistant with a certain user profile presents a scale challenge due to the large number of experiments necessary to cover many commands and user profiles. To address this challenge, we develop an automated system for running our experiments. 
We chose the Alexa voice assistant to build our experimental testbed because it allows us to conduct experiments at scale and to assess both command intelligibility and user voice similarity. In particular, for similarity, the Alexa Skill Developer Test Platform can provide Skills, which have personalization option enabled, a confidence level ranging from 0 to 300. This is not possible with the other voice assistants.

Our system integrates an Amazon Alexa Skill that we developed in our workflow of evaluating the commands generated using concatenative synthesis. Amazon Alexa Skills are third-party applications that add new capabilities to Alexa. By invoking our locally running Skill, we can gain access to a variety of data captured by Amazon Alexa which is processed by the Cloud backend and sent to the Skill. Although other virtual assistants provide similar applications like Google Actions and Samsung Bixby Capsules, they do not allow access to information about confidence level in voice similarity. 
We also take advantage of the Developer Console of Alexa to automate experiment steps ranging  from playing the command to Alexa, to collecting the response for these commands. 

\begin{figure*}
    \centering
    \includegraphics[width = 0.8\textwidth]{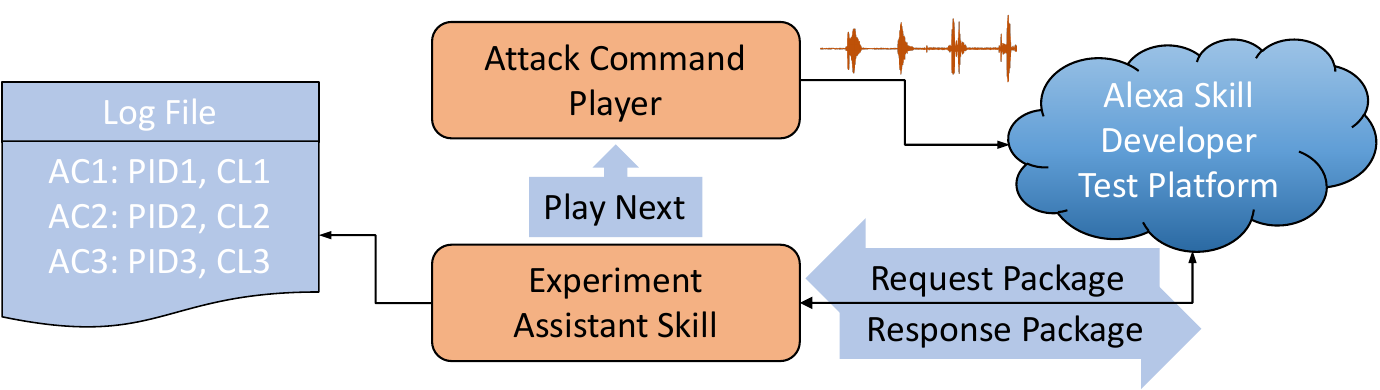}
    \caption{Automated process for evaluating attack efficacy for synthetic commands.}
    \label{fig:auto}
\end{figure*}

As shown in Fig.~\ref{fig:auto}, our Experiment Assistant Skill (EAS) dispatches a message to the attack command player, instructing it to play the next command to Alexa. The command player runs on our local host and has a set of command files. After receiving the message, the player commences the playing of the command on the Alexa Skill Developer Test Platform. It is an online platform provided by Amazon for developers to test their Skills. This platform allows us to interact with Alexa by typing commands or use the browser to record verbal commands through a built-in microphone. We leverage this platform to monitor the packages sent to our EAS and observe the responses. To streamline this process, we utilize Ubuntu's capability to route its audio output back to its input. Consequently, the audio can be delivered internally to the browser without traversing physical speakers and microphones. This enables communication between the attack command player and the web version of Alexa for automated testing purposes.

Alexa performs speech-to-text conversion to determine the command and computes similarity of the command source voice with the user profile to assign a confidence level that the command is uttered by an authorized user. The recognized PersonID (PID) and corresponding Confidence Level (CL) are encapsulated within a JSON file before being transmitted to our EAS. 
The EAS records the results for each Attack Command (AC) and notifies the player to play the next command, while maintaining the Skill session alive by including a keep-alive field in the response package. 
Our automated system allowed us to run large number of experiments for evaluating the efficacy of synthetic attack commands. 

\section{Evaluation}
\label{evaluation}
We evaluate the efficacy of attacks that use commands synthesized from diphones extracted from target speech by conducting a variety of experiments with the testbed described in the previous section. We first list the commands that are used in the experiments and then provide details of the datasets used with the testbed. 

\subsection{Voice Assistant Commands, User Profiles and Command Synthesis}
\subsubsection{Voice Assistant Commands}
Amazon Alexa skills offer a plethora of commands~\cite{shezan2020read} that could be issued via Alexa. Since we want to test commands for many user profiles and when different amount of target speech is available, we chose the following nine commands. Our choice was dictated by length of commands (e.g., number of words) as well as whether the command corresponds to a sensitive action. For example, ``What is my name?'' is less sensitive than ``Asking to unlock a car'' or ``Asking a bank about a transaction''. 
\begin{enumerate}
    \item AC0. Alexa, what is my name?
    \item AC1. Alexa, call my phone.
    \item AC2. Alexa, play some Jazz music!
    \item AC3. Alexa, add bananas to my cart.
    \item AC4. Alexa, when is my dentist's appointment?
    \item AC5. Alexa, ask my bank what my checking account balance is.
    \item AC6. Alexa, ask my bank to make a payment.
    \item AC7. Alexa, ask my bank for my most recent transactions.
    \item AC8. Alexa, ask Chevrolet to unlock my car.
\end{enumerate}

\subsubsection{Setting Up Target User Profiles}
For our experiments, we selected 10 male and 10 female profiles from different geographical locations around the world, as shown in Tables~\ref{tab:male-speakers} and ~\ref{tab:female-speakers} in the Appendix.
Since an Alexa device requires that a user profile be set up prior to issuing commands to use its personalization functionality, we used Coqui TTS VITS model to generate 
the four commands listed below
that Alexa uses to set up a user profile. These four commands were generated for each of the 20 users.
\begin{enumerate}
    \item PC0: Alexa, what's the temperature outside?
    \item PC1: Alexa, play music.
    \item PC2: Alexa, turn off the light.
    \item PC3: Alexa, add milk to my shopping list.
\end{enumerate}

\subsubsection{Experiment Audio Preparation}
The attack commands AC0--AC8, listed above, were generated using the unit-selection concatenative synthesis method described in Section~\ref{section:command_synthesis}.   
For unrelated target speech, we used actual user speech recordings for the chosen profiles in the original VCTK Corpus. The speech in these recordings is unrelated to voice assistant commands because the spoken sentences came from a newspaper, the rainbow passage and an elicitation paragraph used for the speech accent archive.~\cite{veaux2017cstr}
We then extracted diphones from these audio recordings, 
which were used to synthesize all attack commands for each of the 20 profiles.
We also generated the chosen command audios with Coqui TTS using \verb|vits_vctk_fine_tuned| model for all 20 profiles selected by us. These command audios provide a baseline for our experiments.
\subsection{Limited Speech and Diphone Coverage}
\label{diphone_coverage}

\begin{figure}[htbp]
  \centering
{\includegraphics[width=0.45\textwidth]{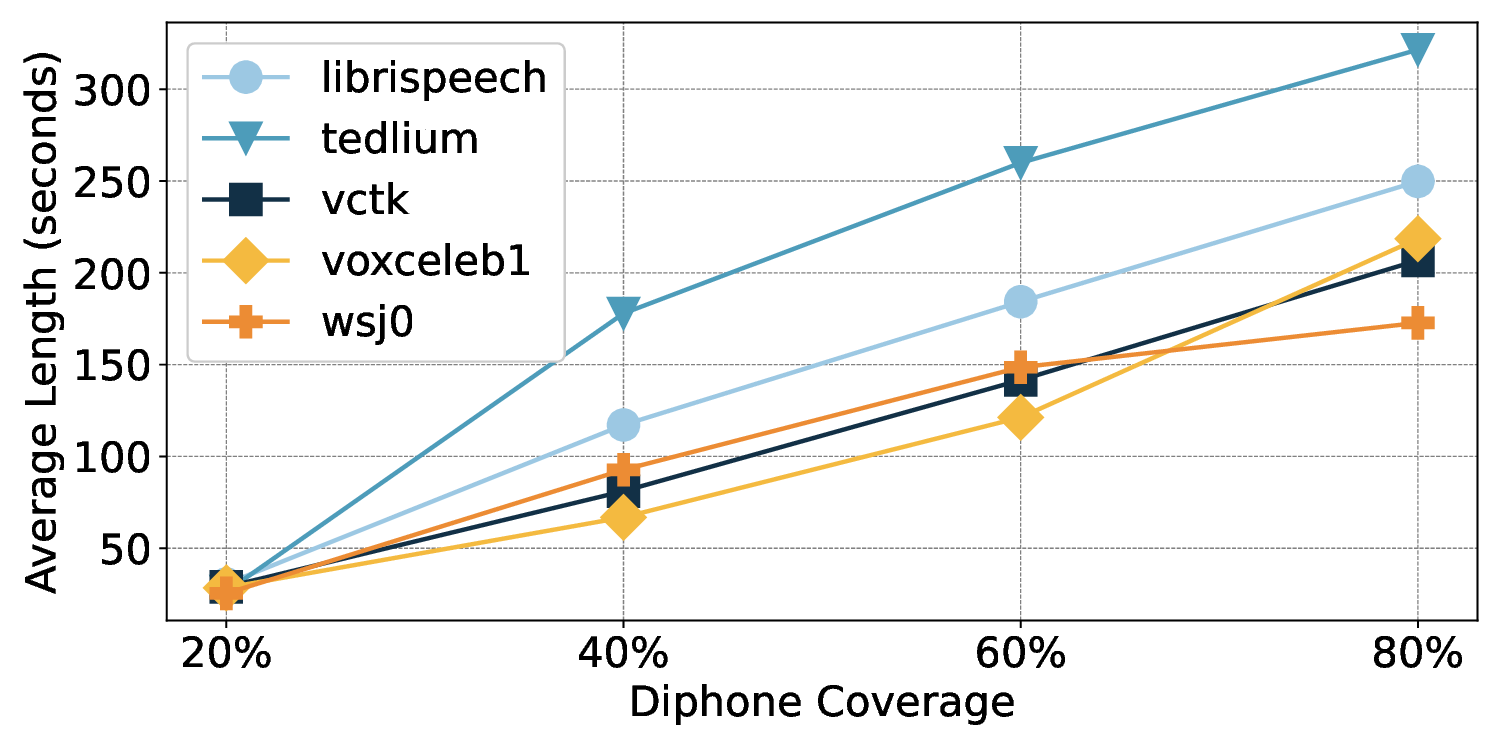}}
  \caption{Average Audio Lengths in Seconds for Different Coverage. (The length is estimated assuming there are 750 diphones in one minute speech. )}
  \label{fig:length_coverage}
\end{figure}

Our threat model assumes limited and unrelated target speech which may not have all the diphones needed for the synthesis of a command. This raises two questions. First, how does the coverage of diphones depend on the amount of available speech. Second, if only limited amount of speech is available, what diphones are likely to be present and which ones may be missing. In the general case, answers to these questions are challenging because they depend on the specific speech that is available for a target. We explore their answers using several popular datasets. Table \ref{tab:dataset_overview} provides an overview of these datasets, detailing the type of speech source and the main content of the audios. These datasets span a range of genres, from audiobooks to TED talks to celebrity interviews, providing a broad view of diphone distribution across different types of spoken content. We first show the relationship between diphone coverage and the length of the speech in these datasets. The coverage computation is based on diphones required to synthesize all nine attack commands AC0--AC8. Some of the datasets have both speech and transcripts and others only have speech. For these, we first use a speech-to-text system to generate transcripts. The transcripts are used to extract diphones and correlate a diphone sequence with the length of speech from which it is derived. Figure~\ref{fig:length_coverage} depicts how diphone coverage increases with the length of speech. As can be seen, 50\% diphone coverage can be achieved with speech ranging from 2 to 4 minutes. On the other hand, 80\% coverage could require up to five minutes of speech.

We computed frequencies of diphones across all speech samples in our datasets to answer the the second question which addresses the diphones that are more likely be present when only limited amount of speech is available. More specifically, if diphone coverage in available speech is $p\%$, we assume that the more frequent diphones will likely be available. Thus, if a command required a certain set of diphones but the coverage is only 50\%, only half the required diphones are chosen from speech available from the target. In our experiments, we chose diphones based on their frequency or popularity computed from our datasets. For example, in case of 50\% coverage, the diphones are ordered by popularity and the top half are chosen. We believe this is reasonable because diphone frequency across the diverse datasets should capture diphone distribution in unrelated speech collected for attack targets.

\begin{table}
\centering

\begin{tabular}{lll}
\toprule
\textbf{Dataset} & \textbf{Audio Source} & \textbf{Main Content}  \\ \midrule
LibriSpeech & Audiobooks & Various books  \\ 
VoxCeleb1 & YouTube & Celebrity interviews  \\ 
VCTK & Read Speech & Phonetically balanced sentences \\ 
WSJ0 & News Text & Wall Street Journal news articles  \\ 
TED-LIUM & TED Talks & Transcriptions of TED Talks  \\ 
\bottomrule
\end{tabular}
\caption{Overview of the Datasets Used in This Study.}
\label{tab:dataset_overview}
\end{table}

\subsection{Experiments}
The results of each experiment that we run are captured by the Skill that we developed and deployed on a server locally on our machine running Ubuntu LTS 20.04. Experiments were conducted by varying commands, the command synthesis method and the fraction of diphones available for the synthesis of a command. In each run, the Skill receives information about the command intent (e.g., what command is being issued to the voice assistant), and an identifier of the user who has issued the command. A confidence level is also received which captures the similarity between the received command's issuer and the user whose identifier is provided.

We define success of attacks launched with synthetic attack commands based on two metrics. First, {\it intelligibility} captures if the command intent received by our Skill matches the issued command. Second, {\it similarity} captures the confidence level that the command is issued by the user matching the user identifier received by the Skill. Skills can use such confidence levels for speaker match when using voice biometric authentication to decide if a command should be processed. Currently, only four confidence levels are returned (300, 200, 100, 0), where 300 indicates the highest voice similarity.
Although our analysis is based on a black-box approach, we could gain access to the command audio transcript which likely is used for such intent detection. A command may not be correctly identified when synthesized speech quality is poor and the transcript from which command intent is inferred has many word errors. To further explore when intents were not correctly identified, we used a {\it word error rate} (WER) metric in the command audio transcript. A word error results in a command transcript when a word is changed, new word is inserted or a correct work is dropped. WER is defined by the ratio of number of erroneous words divided by the total number of words in the command.

\subsection{Results}

We conducted experiments to measure both command intelligibility and target profile similarity and present their results next.

\subsubsection{Command Intelligibility}
Once a profile is set up with natural sounding speech, we issue attack commands AC0--AC8 that are generated using our concatenative synthesis method where diphones are extracted from actual speech audios from VCTK Corpus.
Each synthesized command is played using our testbed and our Skill is used to find the command intent. If the intent matches the played command, that denotes success. Otherwise, the command is not correctly understood and not executed. Thus, the fraction of commands where the intent is correct is the success rate. 

\begin{figure*}[htbp]
  \centering
  \subfloat[Total Correct Intent Count for Each Male Profile.]{\includegraphics[width=0.44\textwidth]{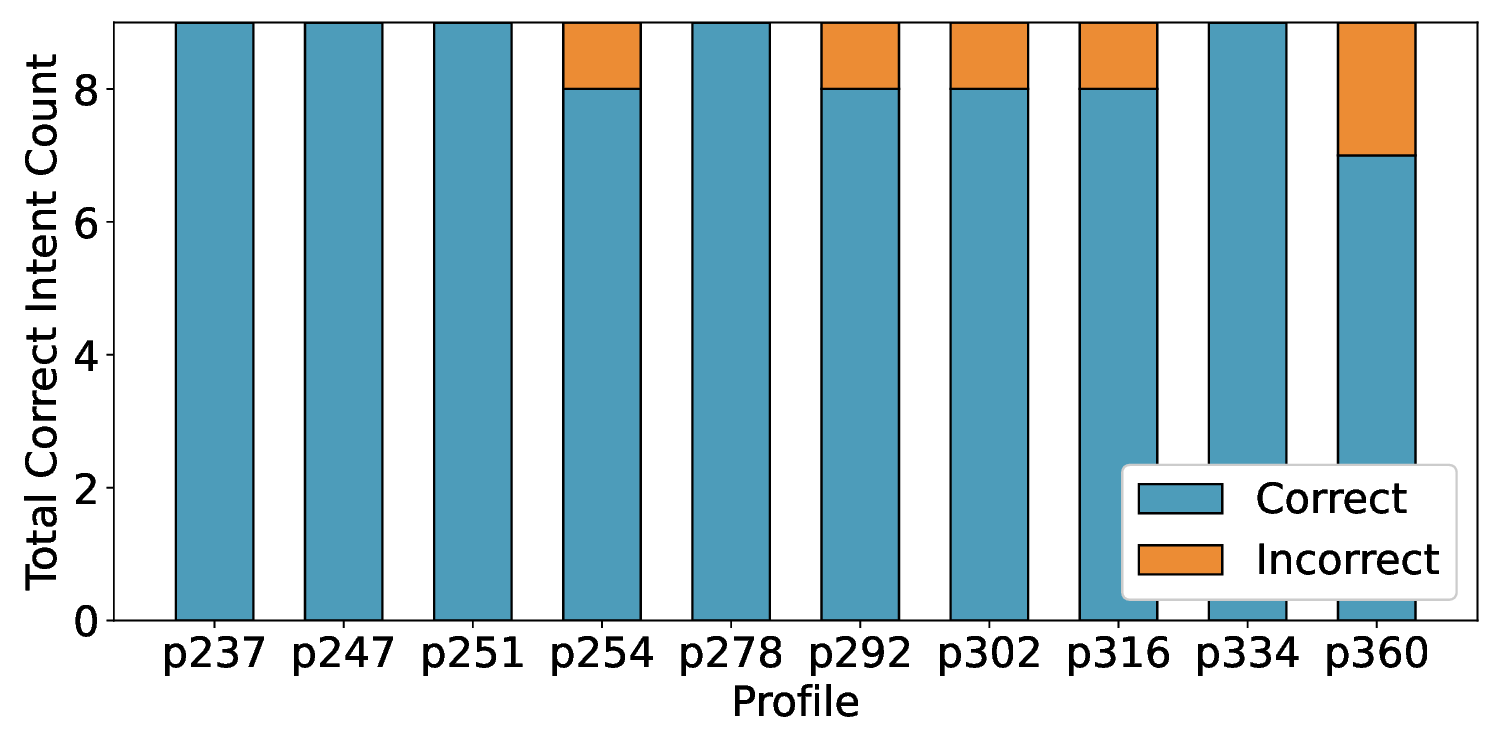}\label{fig:male_intent}}
  \hfill   
  \subfloat[Total Correct Intent Count for Each Female Profile.]
  {\includegraphics[width=0.44\textwidth]{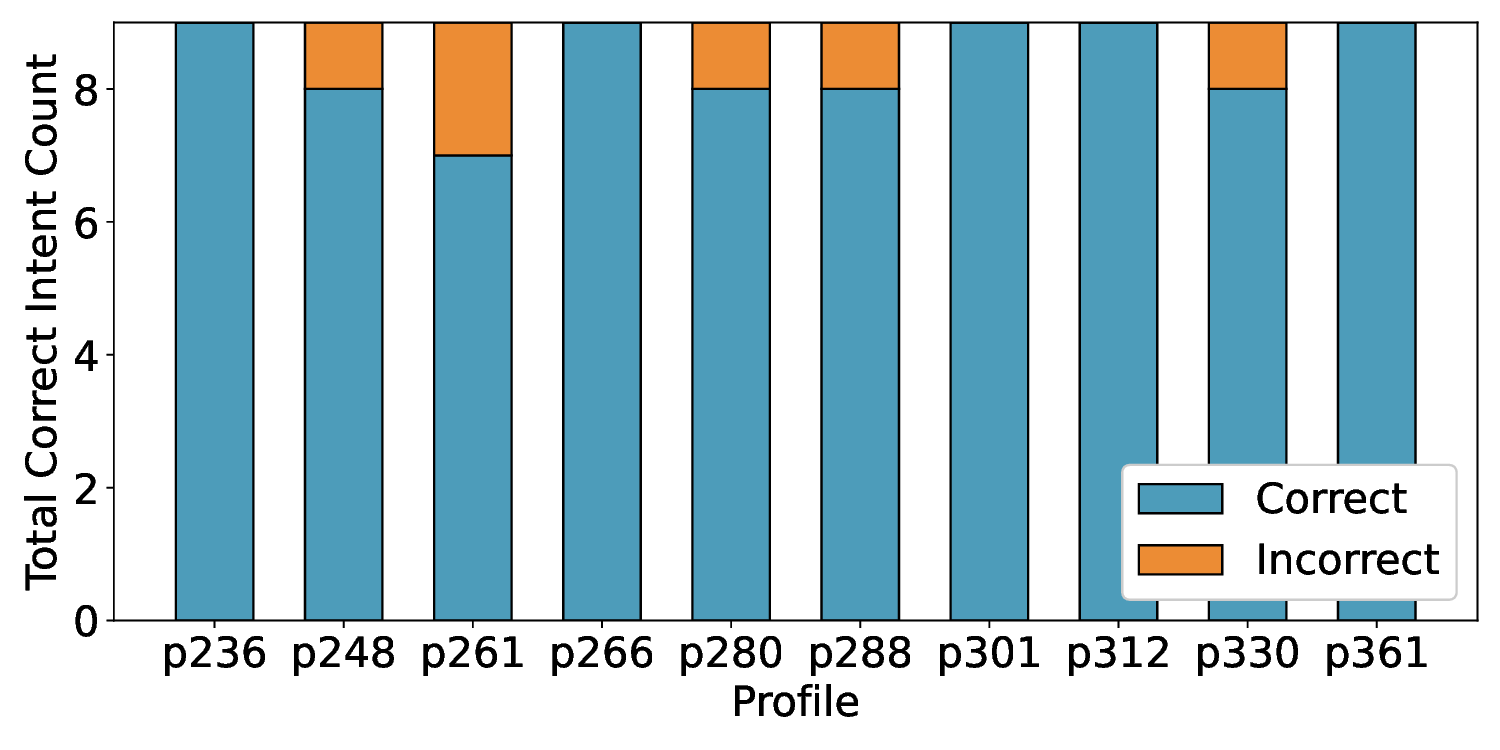}\label{fig:female_intent}}
  \vspace{\baselineskip} 
  \subfloat[Total Correct Intent Count for Each Command.]{\includegraphics[width=0.44\textwidth]{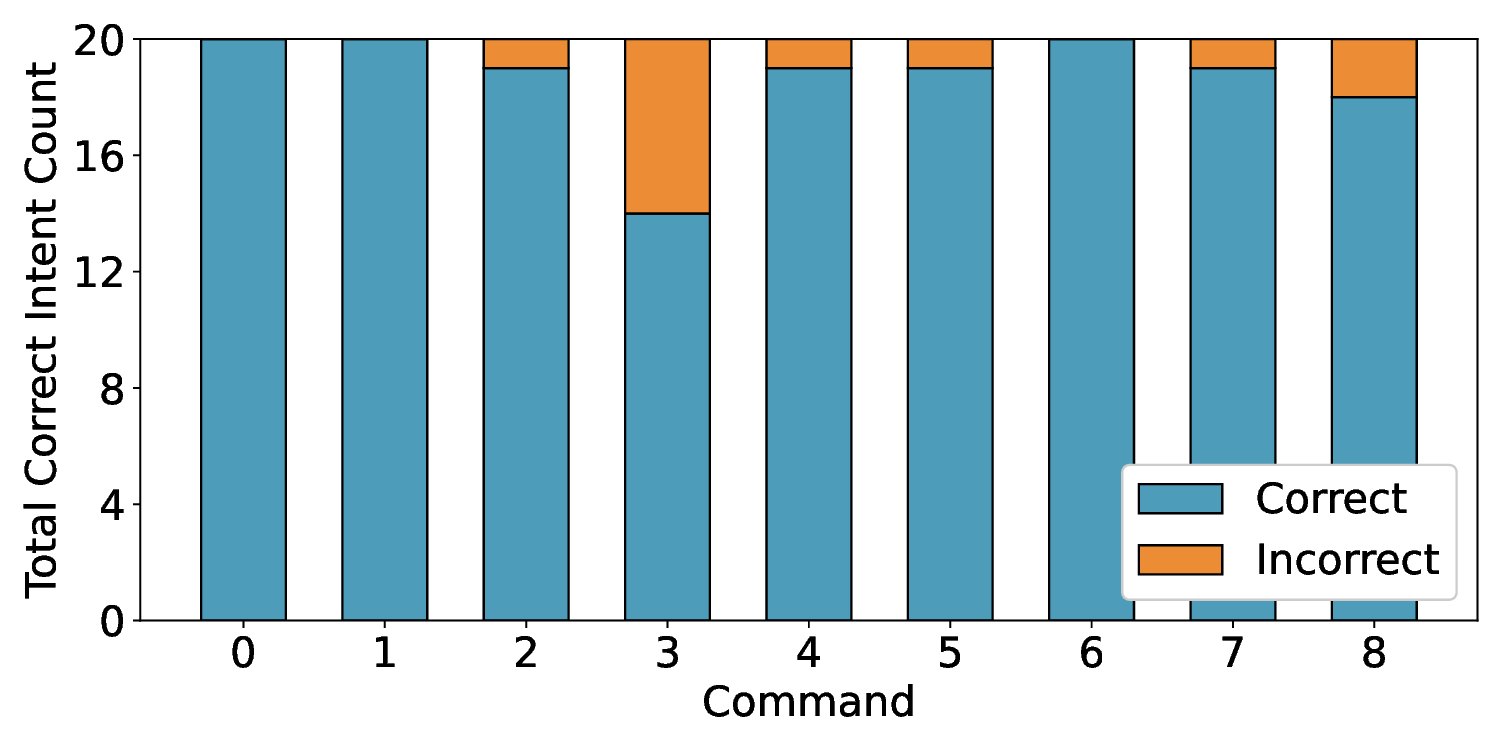}\label{fig:command_intent}}
  \hfill
  \subfloat[Total Correct Intent Count for Different Diphone Coverage of Short Command and Long Command.]{\includegraphics[width=0.44\textwidth]{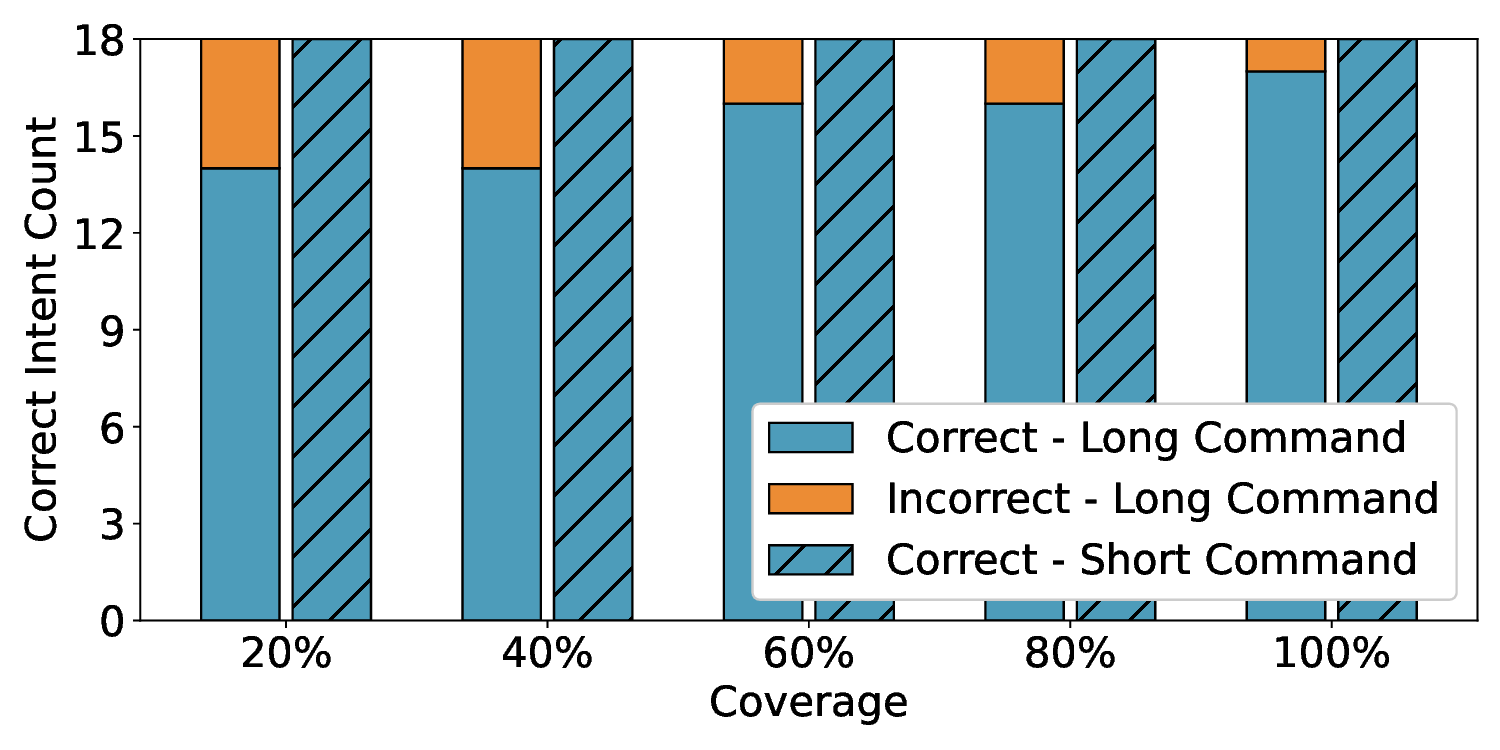}\label{fig:intent_partial}}

  \caption{Synthesized Command intelligibility analysis.}
  \label{fig:intel}
\end{figure*}

We tested nine commands for 20 different user profiles (10 male and 10 female).
Fig.~\ref{fig:intel} shows the results of our experiments. In Fig.~\ref{fig:male_intent} and~\ref{fig:female_intent}, the x-axis shows the different user profiles and y-axis shows the number of commands for which the intent is correctly recognized. Fig.~\ref{fig:male_intent} and~\ref{fig:female_intent} show results for male and female user profiles separately. Fig.~\ref{fig:male_intent} shows that for five of the ten male user profiles, intent for all 9 commands are correctly identified. For the other four profiles, the intent for only one command is missed. Thus, across 10 male profiles and 9 commands, 84 of the 90 command intents are correctly recognized. The results are similar in Fig.~\ref{fig:female_intent}.
Thus, our results show that a low cost concatenative synthesis method is able to produce commands for both male and female users with high intent accuracy. 
Overall, as seen in Fig.~\ref{fig:command_intent}, intent for three commands is correctly recognized across all 20 user profiles. We observe that intent for longer commands that have more words (Commands AC2 - AC8) are missed a few times. AC3 is missed the most which could be because the diphones used for the word "banana" do not match well with its required diphones.
Across all profiles and commands, the intent is correctly recognized in 169 of the 180 tests, resulting in 93.8\% intelligibility. This shows the efficacy of the unit-selection based command synthesis method for generating attack commands even when they do not sound natural to the human ear. 

If only limited amount of speech is available for a targeted user, the extracted diphones from it may not provide a full coverage of the diphones needed to synthesize a command. In this case, missing diphones can be replaced by the same diphones extracted from speech from another user of the same gender. For conducting experiments with partial diphone coverage, one question that arises is how to choose the diphones that are available for the targeted user. As discussed in Section~\ref{diphone_coverage}, in our experiments, we preferentially choose diphones based on their frequency in the speech in a diverse set of datasets analyzed by us. 

Fig.~\ref{fig:intent_partial} shows that intent recognition is resilient to missing diphones for a target for a short command (Command AC0) while longer commands (e.g., Command AC5) being more impacted by them. The missing diphones were substituted with diphones from profile p288 for female and profile p360 for male targets as they provide the highest average confidence level in cross attacking experiments shown in Fig~\ref{fig:cross}. We vary diphone coverage of each profile (Except for p288 and p360) from low (20\%) to full coverage for Command AC0 and Command AC5. All intents for the Command AC0 are correctly determined at all coverage levels. For Command AC5, which had the most missed intents with 20\% diphone coverage, the missed intents decreased as the coverage grew.
%
%
%

\begin{figure}[!ht]
  \centering
  \subfloat[Average WER for Each Command.]{\includegraphics[width=0.45\textwidth]{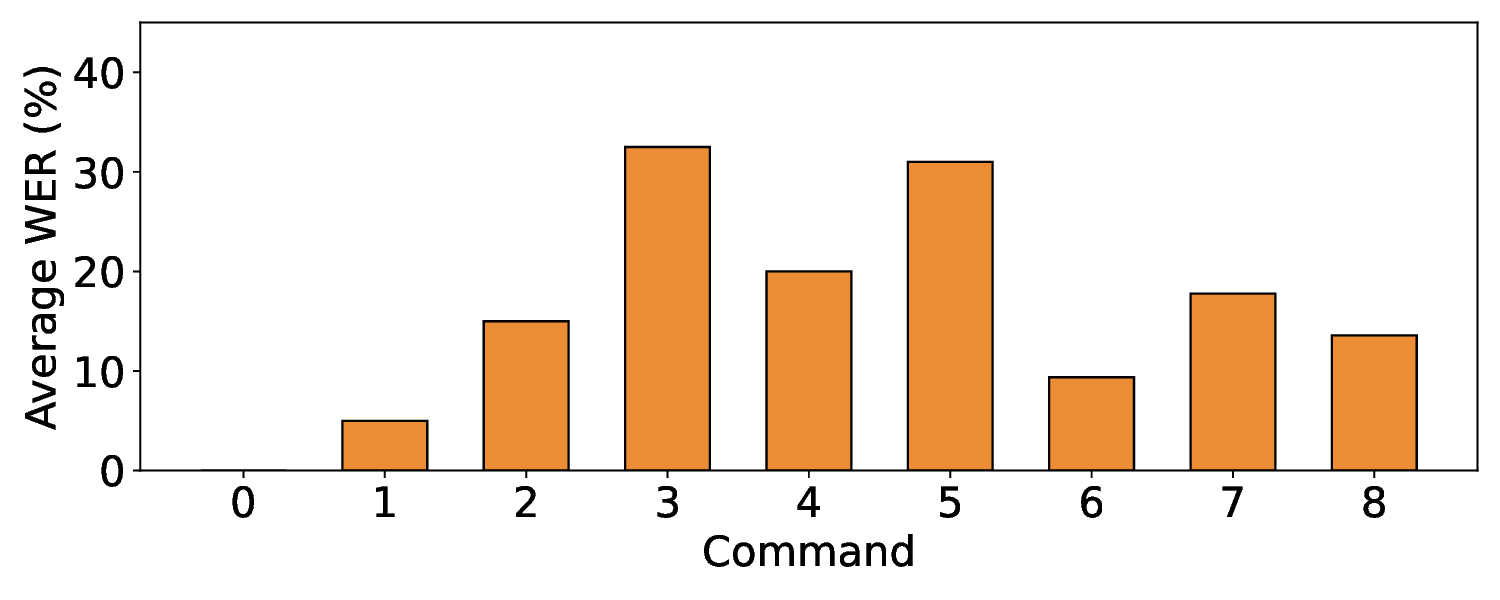}\label{fig:command_wer}}
  \vspace{\baselineskip} 
  \subfloat[Average WER for Various Diphone Coverage of Commands.]{\includegraphics[width=0.45\textwidth]{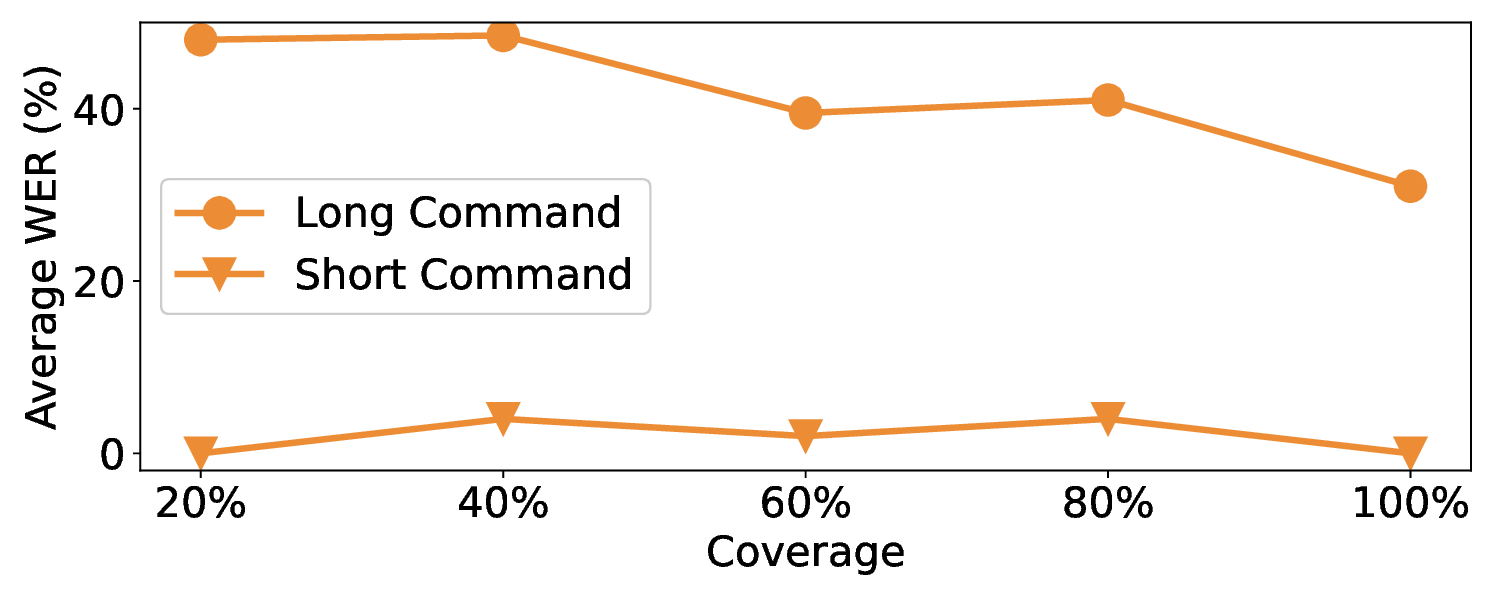}\label{fig:wer_partial}}

  \caption{Word Error Rate in Command Transcripts.}
  \label{fig:error}
\end{figure}

{\bf Word Error Rate (WER) Analysis:}
A command intent is inferred by Alexa from the transcript generated from the command audio.~\cite{alexa_developer_guide}
To further explore why the command intent is not inferred correctly in a small number of cases, we examine the transcripts for the various commands 
and calculate the average Word Error Rate (WER) for each profile and command, as shown in Fig.~\ref{fig:command_wer}. The average WERs for most commands are below 20\%, which falls within acceptable WER thresholds \cite{Microsoft2023}. The exceptions are commands AC3 and AC5. We found that Command AC3 is hard is because the word "banana" sometimes is confused with other words, and Command AC5 is the longest command, comprising of 9 words.


A natural question is if WER will increase with lower diphone coverage.
In Fig.~\ref{fig:wer_partial}, the x-axis represents different coverage of diphones, while the y-axis shows the corresponding WER. It can be seen that the coverage of diphones does not have much impact on WER for short commands. However, as shown in Fig.~\ref{fig:intent_partial}, we observe that despite fluctuations in WER within a certain range, it has little influence on the correctness of intent recognition. 
While it is important to consider WER, it is worth noting that despite some errors, Amazon Alexa is capable of correctly identifying the intent of the command in most cases.
In a large-scale attack scenario where many users are targeted, we can expect a high rate of successful intent recognition for most of the attack commands.

\subsubsection{Profile Similarity Analysis}
For an attack to be successful, not only the command must be recognized correctly but the voice assistant must determine that the command is coming from an authorized user of the voice assistant. 
Although Alexa does not claim to perform voice-biometric authentication, it does compare how similar the voice of the source of a command is to the users for whom profiles exist. Skills are provided a UserId for the source of a command and a confidence level in the similarity of the voice of the command source and the user having the returned UserId. In the absence of additional authentication evidence (e.g., a PIN), the highest returned confidence level is 300, which suggests a close match. Other confidence level values observed by us include 200 (medium), 100 (low)  and 0 (very low or no match). The Skill deployed by us allowed us to assess the confidence level with the UserId information to evaluate the similarity between commands synthesized with our unit-selection method and with the natural sounding commands generated with the well trained Coqui TTS model.

We designed experiments to answer two questions: (i) how does confidence level change when a command is generated with the concatenative synthesis method compared to when the same command is generated with Coqui, and (ii) how distinguishing confidence values are when a command is issued by a user that does not have a match with the profile set up on Alexa. To answer these questions, we set up Alexa target profiles for 20 different users (10 male, 10 female). For each of these target profiles, we issued commands synthesized with the two methods for all of the user profiles. 
Since diphone synthesized commands are played when command source matches the target profile for all 20 users, this will help us answer the first question. By playing commands for users having profiles that do not match the target, we are able to answer the second question. We next discuss the results of these experiments.

\begin{figure*}[!ht]
  \centering
  \subfloat[Female Profiles Cross Attacking. Commands generated from Coqui TTS.]{\includegraphics[width=0.31\textwidth]{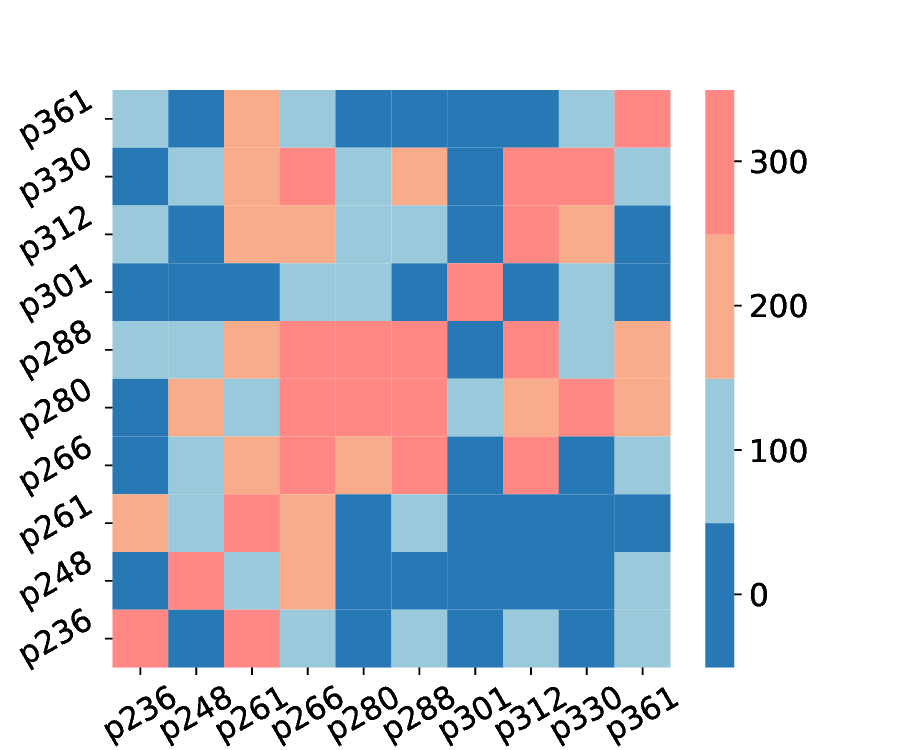}\label{fig:female_tts}}
  \hfill
  \subfloat[Female Profiles Cross Attacking. Commands generated from Unit-Selection Synthesis.]{\includegraphics[width=0.31\textwidth]{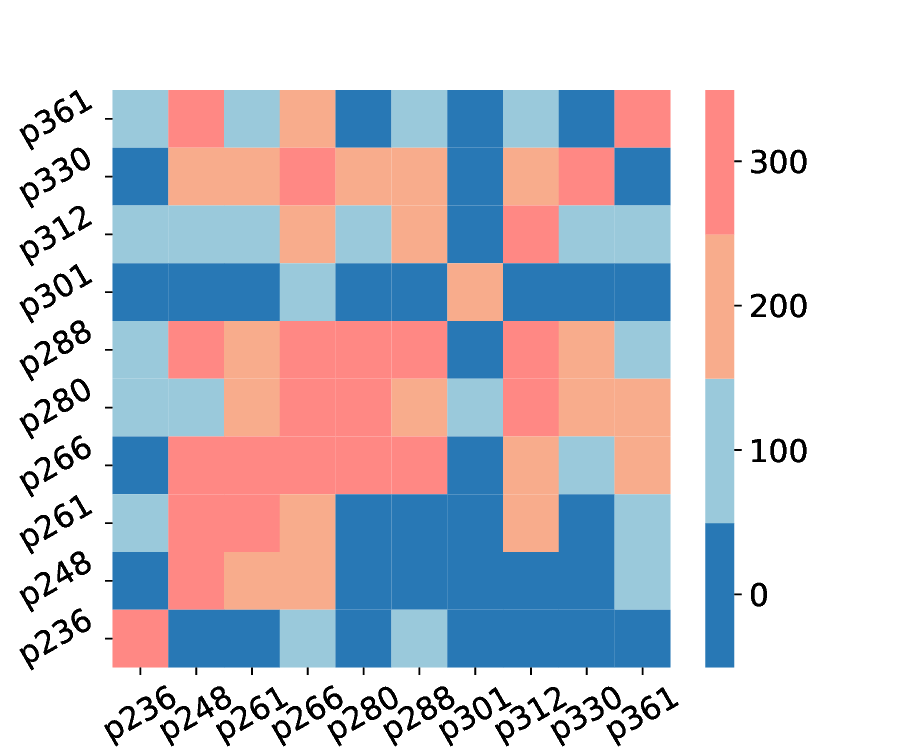}\label{fig:female_diphone}}
  \hfill
  \subfloat[Difference between (a) and (b)]{\includegraphics[width=0.31\textwidth]{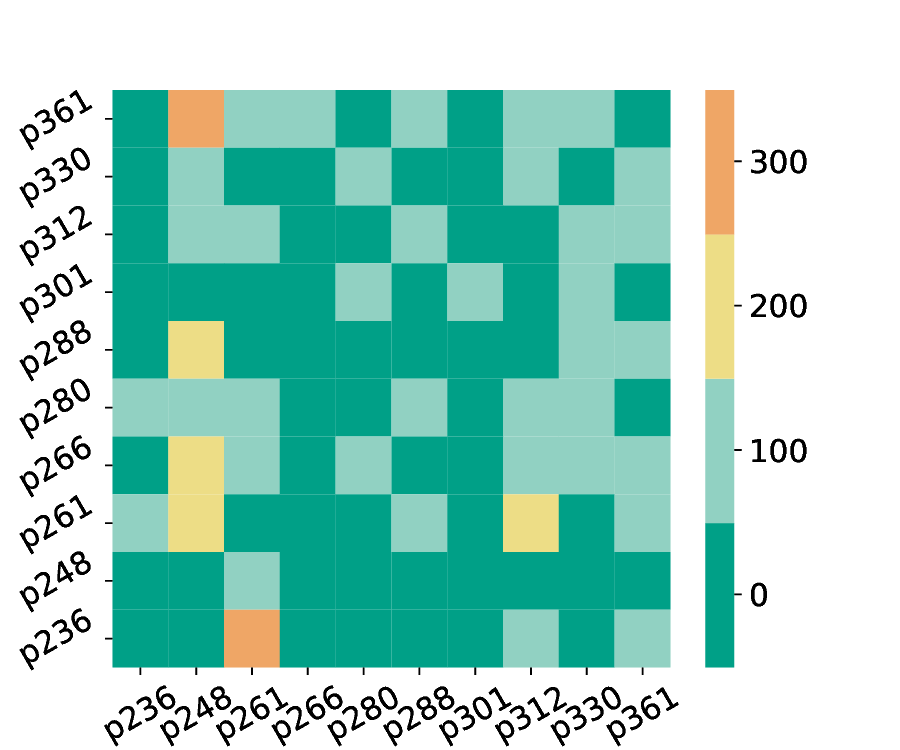}\label{fig:female_difference}}
  \vspace{\baselineskip} 
  \subfloat[Male Profiles Cross Attacking. Commands generated from Coqui TTS.]{\includegraphics[width=0.31\textwidth]{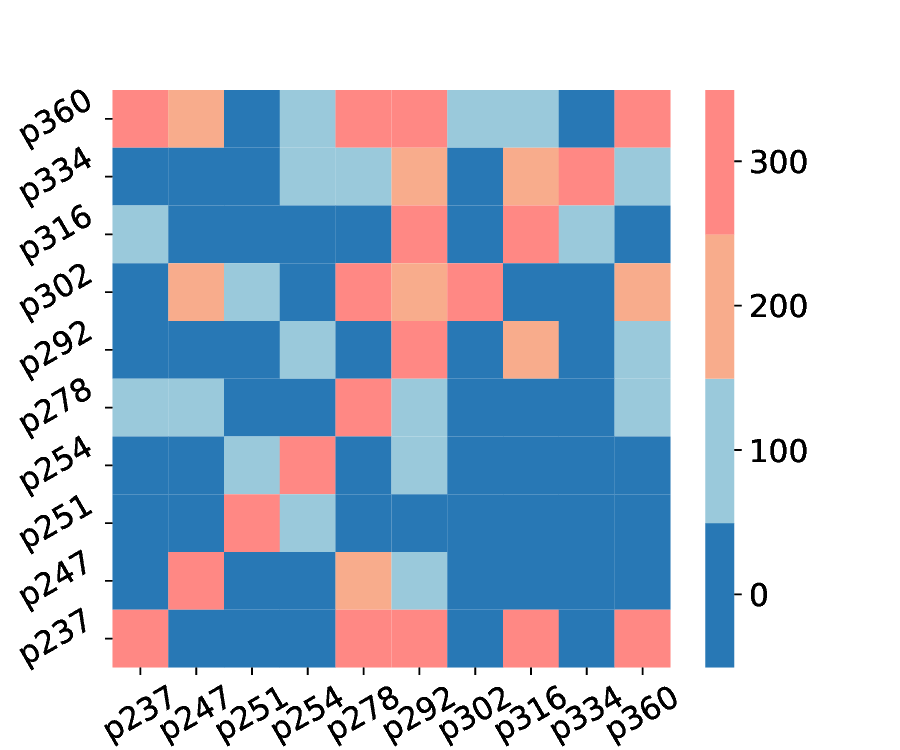}\label{fig:male_tts}}
  \hfill
  \subfloat[Male Profiles Cross Attacking. Commands generated from Unit-Selection Synthesis.]{\includegraphics[width=0.31\textwidth]{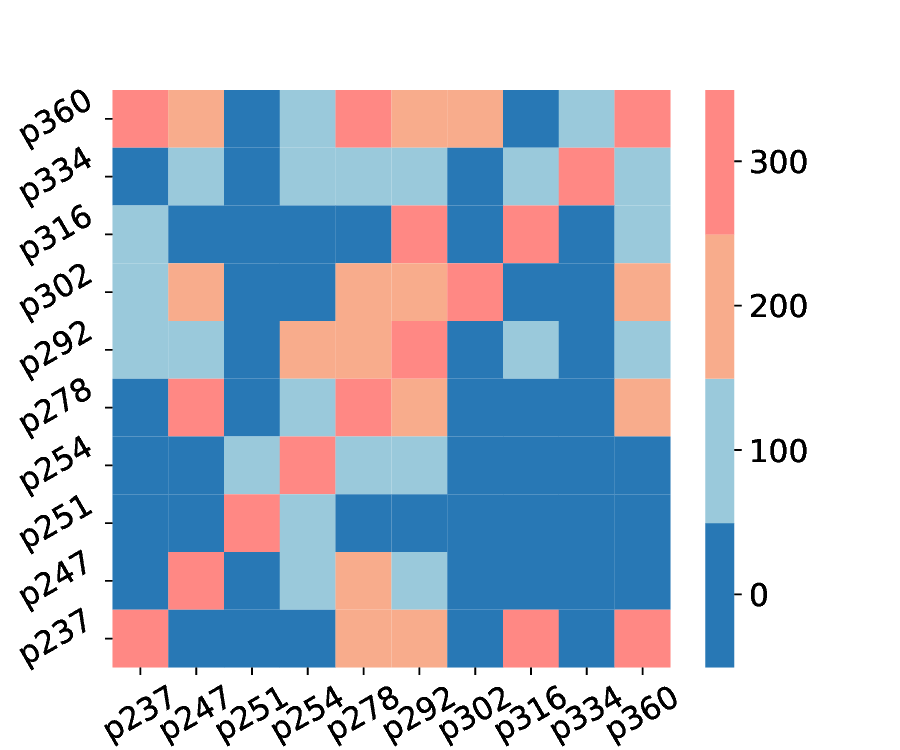}\label{fig:male_diphone}}
  \hfill
  \subfloat[Difference between (d) and (e).]{\includegraphics[width=0.31\textwidth]{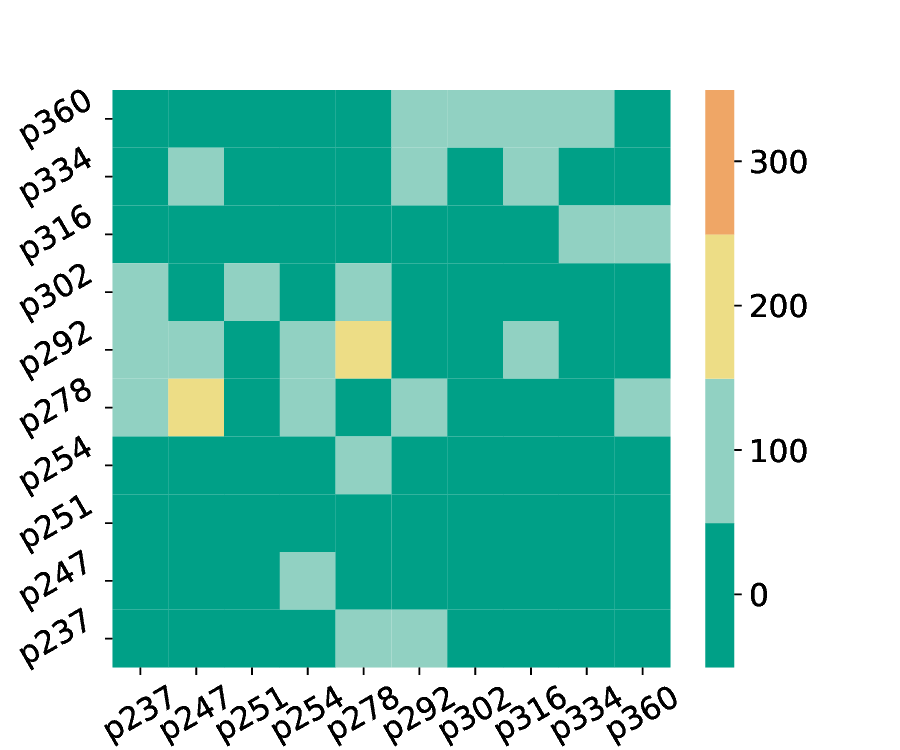}\label{fig:male_difference}}
  \caption{Profile Similarity Analysis.}
  \label{fig:cross}
\end{figure*}

The results of our experiments are shown in Fig.~\ref{fig:cross}. We present them as a matrix with color coded entries. Both axis are labeled with each of the tested  profiles and entry ($p_i$,$p_j$) denotes the confidence level achieved when Alexa is setup with target profile $p_i$ and the synthesized command that is directed to it comes from speech of user profile $p_j$. The confidence level values are shown by the color of the entry (300 is pink and 0 is shown in blue). Our results are shown for both female and male profiles. Fig.~\ref{fig:female_tts} shows it for female profiles when the attack commands are generated with Coqui TTS and Fig.~\ref{fig:female_diphone} shows the same when the commands are synthesized with the unit-selection method. Fig.~\ref{fig:male_tts}~and~\ref{fig:male_diphone} show the same results for male profiles.
Our results show that when the target profile on Alexa matches the profile which is the source of an attack command, the highest confidence level is returned independent of the method used for synthesizing the command. In fact, the unit-selection method achieves the same high confidence value 300 in 19 of the 20 user profiles. For one female profile, the confidence level for unit-selection command is 200 compared to 300 for Coqui TTS. These results demonstrate that compared to the more natural speech produced by Coqui TTS, unit-selection suffers minimal or no degradation in profile similarity. 

To check that confidence levels do depend on similarity of a command source's voice and the voice used to setup an Alexa profile, we conducted experiments where we set up an Alexa target profile with one user's voice and issued the command in the voice of a different user. We conducted such experiments separately for male and female profiles. Our results are summarized in Fig.~\ref{fig:female_tts},~\ref{fig:female_diphone},~\ref{fig:male_tts} and~\ref{fig:male_diphone}. We can see that confidence level does degrade when command source does not match the user for whom an Alexa profile is set up. For Coqui TTS, the confidence level between mismatched command source and Alexa profile is below or
equal to 100 in 136 of the 180 pairs. This same number is 127 for the unit-selection method synthesized commands. Thus, commands synthesized with both methods have low confidence levels for non-matching users and the confidence level is 100 or lower for a majority of the pairs.

More interestingly, differences in confidence level do not depend on the command synthesis method. We got similar confidence levels for commands generated with Coqui TTS and unit-selection methods. This is visually shown in Fig.~\ref{fig:female_difference} and~\ref{fig:male_difference}. For 180 cases across all profiles for both genders, when the target profile is different from the command source profile, we have exact match for the two methods for 112, and the confidence is different by only one level in 60 cases. Only in 8 cases the difference is 200. 

\subsubsection{Confidence Level Analysis for Different Diphone Coverage}
\begin{figure}[htbp]
  \centering
  {\includegraphics[width=0.45\textwidth]{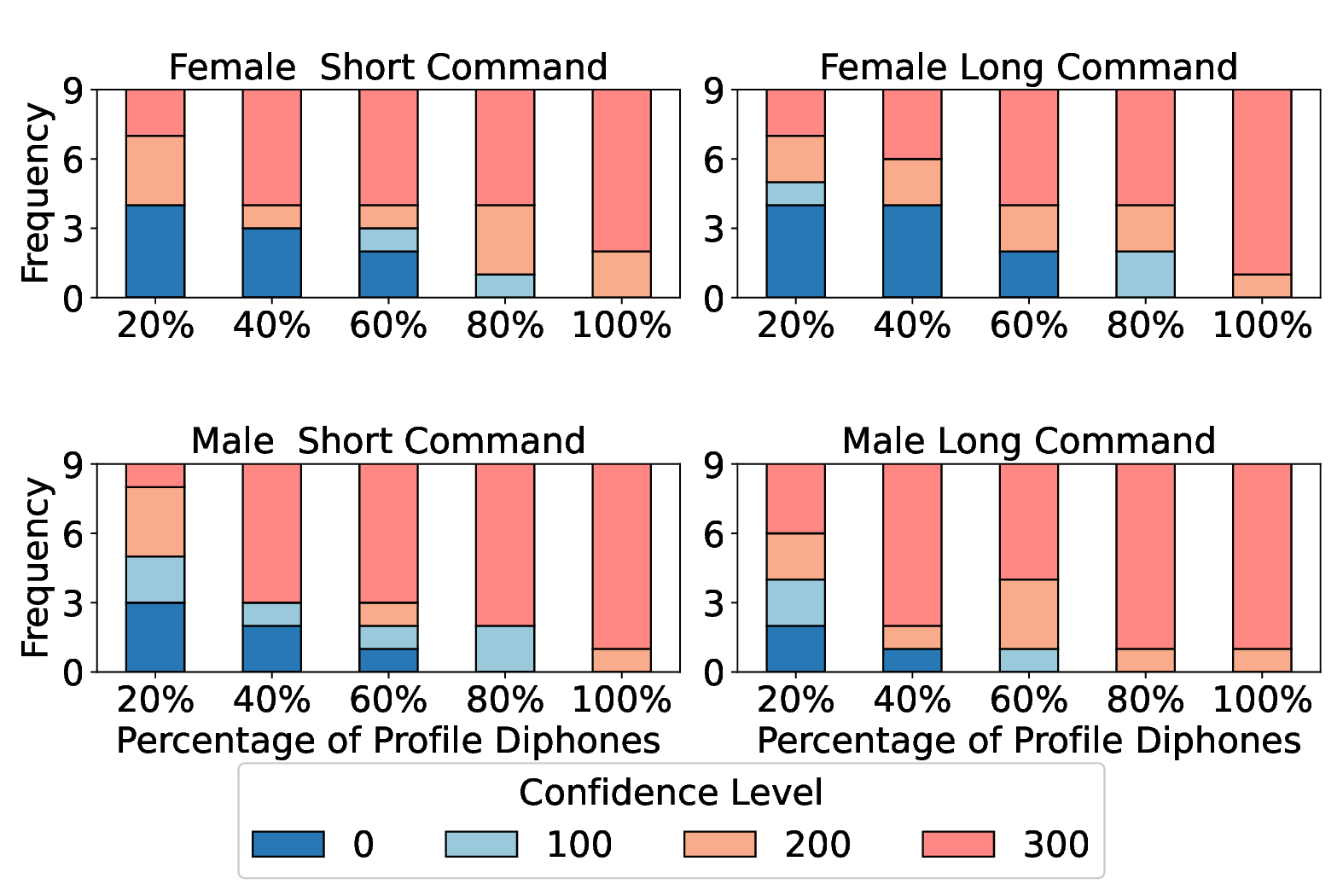}}
  \caption{Diphone Percentage and Confidence Level.}
  \label{fig:sorted_partial}
\end{figure}

In case when only a limited amount of speech is available for a target, similar to our analysis for intent detection, we utilize stored diphones from another user to substitute the missing ones in our unit-selection method. In our experiments, we used diphones from profile p288 for female and profile p360 for male targets as they provide the highest average confidence level in Fig.~\ref{fig:cross} when they are used to synthesize attack commands for the profiles of each gender. 
We conduct experiments to show how confidence level varies for a short command (Command AC0) and a long command (Command AC5) when diphone coverage of the command source ranges from 20\% to 100\%. 
As depicted in Fig.~\ref{fig:sorted_partial}, the distribution of confidence levels for each coverage level is visually presented, with a larger area indicating a higher frequency of corresponding confidence levels. Notably, we observe that higher coverage percentages tend to correlate with an increased likelihood of achieving confidence levels of 300. Also, male profiles tend to have higher confidence level when comparing with female profiles.
The results suggest that even with a relatively modest 20\% coverage of the target's diphones, the likelihood of successfully passing the speaker match process remains about 50\% with synthetic commands. This observation is consistent with our findings in Fig.~\ref{fig:cross}, where we noted that the best-performing profile exhibited a high success rate even when it lacked any audio samples from the target.

\subsubsection{Attack Footprint}
A number of options exist in launching an at scale attack that targets a large number of voice assistants that are commanded by malware controlled compromised devices like nearby computers. 
A stealthier method for such malware is to utilize local resources at the infected computers to synthesize the attack commands. To do this, the malware must download code and data required for command synthesis. In this section, we discuss 
how the diphone based concatenative synthesis method enables a significantly smaller network and host footprint for such malware which can help evade detection. We show this by comparing the resources needed by a high quality synthesis model such as Coqui TTS with those required by the concatenative synthesis method. Since we assume all models run locally on the target's computer, network traffic includes downloading of both code and data. We pack the code for each method into an executable program using \verb|pyinstaller| to determine code size. 
For measuring the footprints when running, we synthesized the same set of nine commands with both methods, all for the same profile, p236. We extracted diphones from audios with length 180 seconds, as it can cover 40\% of diphones as seen from Fig.~\ref{fig:length_coverage}. To minimize interference from other processes, we executed each model separately. Our experiments were conducted on a computer equipped with an eight-core processor and 64GB of memory. During the execution of each command synthesis method, we continuously sampled process information using the Linux \verb|top| command at one-second intervals.

As shown in Table~\ref{table:comparison}, resource utilization of TTS, which is representative of high quality speech synthesis, is significantly higher than the concatenative synthesis method. The code and data required by TTS is 158 MB, compared to 34 MB required by our light-weight concatenative synthesis including the extraction libraries, \textit{MFA}. 
CPU usage analysis showed that Coqui TTS uses significantly higher computational resources, averaging 224\%, with peaks reaching 800\% CPU utilization. In contrast, unit-selection averaged at 98\%, with peaks at 100\%. 
Similarly, Coqui TTS memory utilization averaged 491 MB, with peaks at 975 MB, whereas unit-selection used less, averaging 109 MB, with peaks at 184 MB.
These  experiments demonstrate that concatenative synthesis used by us not only is effective in generating attack commands successfully, but it has a much smaller network and host resource footprint.

\begin{table*}[htbp]
\centering
\begin{tabular}{cccc}
\toprule
\textbf{Metric} & \textbf{TTS } & \textbf{Diphone Extraction} & \textbf{Unit-Selection}\\
\midrule
Program + Data (MB) & 7 + 151 & 13 + 0 & 11 + 10 \\
Time (s) & 31 & 32 & 8 \\
Average CPU (\%) & 224 & 66 & 98 \\
Max CPU (\%) & 800 & 100 & 100 \\
Average Memory (MB) & 491 & 164 & 109 \\
Max Memory (MB) & 975 & 221 & 184 \\
\bottomrule
\end{tabular}
\caption{Computing Resources: TTS vs Unit-Selection.}
\label{table:comparison}
\end{table*}

\section{Limitations}
\label{discussion}
Our research findings offer insights into the security vulnerabilities of current voice assistant technologies and the potential risks they could pose. 
However, there are several limitations of our work. Our experiments were conducted with the Amazon Skill Developer Test Platform which allowed us automate running of the experiments and collect data for many profiles and commands. It also allowed us to explore both command intelligibility and speaker voice similarity. We believe our results should be applicable for other voice assistants but we have not conducted similar experiments with them.

Voice assistant applications that perform sensitive actions will likely require high confidence in assessing if a command comes from an authorized user. Our results show that for high confidence, target speech that is several minutes long will be required for sufficient diphone coverage. Thus, the unit-selection method cannot be used when only a small amount of target speech is available (e.g., recording of a short phone call). Other more sophisticated speech synthesis techniques can overcome this limitation but have not been explored by us.

Our threat model relies on an attacker compromising a device close to a voice assistant and using its microphone, speaker and computational resources to launch an attack. If the user of such a device is present when an attack is launched, the command delivered by the device speaker could be heard by the user. We do not address how such commands can be made inaudible or stealthy which has been addressed by other work~\cite{diao2014your, jang2014a11y, Carlini:2016vl, schonherr2018adversarial, Carlini:2018wj, Abdullah:2018ho, Yuan:2018um, kumar2018skill, qin2019imperceptible}.


\section{Conclusion}
\label{conclusions}
Voice is easy to acquire for a large number of targets, raising concerns about large scale attacks with voice assistants like the ones that come via the email or web channels. We demonstrated that limited and unrelated speech of a target can be used with a relatively simple concatenative synthesis techniques to generate attack commands. Such synthesized commands can be processed by current voice assistants like Alexa even when Skills check similarity of command voice with the voice of authorized users who setup profiles on such devices. The testbed developed by us allowed us to show that this holds for a variety of commands for a diverse group of users.
Our research approach, despite its limitations, still demonstrates the need for developing effective defenses against malicious commands that could target voice assistants. Such defenses face many challenges and need to be explored by future research. They need to balance the ease of use that current voice assistants offer with increased security that future voice-enabled applications will demand.

\bibliographystyle{ACM-Reference-Format}
\bibliography{ref}

\pagebreak
\appendix
\section{Appendix}

\begin{table}[h]
\centering

\begin{tabular}{cccc}
    \toprule
\textbf{ID} & \textbf{AGE} & \textbf{ACCENTS} & \textbf{REGION} \\
    \midrule
p237 & 22 & Scottish & Fife \\
p247 & 22 & Scottish & Argyll \\
p251 & 26 & Indian &  \\
p254 & 21 & English & Surrey \\
p278 & 22 & English & Cheshire \\
p292 & 23 & NorthernIrish & Belfast \\
p302 & 20 & Canadian & Montreal \\
p316 & 20 & Canadian & Alberta \\
p334 & 18 & American & Chicago \\
p360 & 19 & American & New Jersey \\
    \bottomrule
\end{tabular}
\caption{\label{tab:male-speakers}Male Speakers.}
\vspace{6px}

\centering

\begin{tabular}{cccc}
    \toprule
\textbf{ID} & \textbf{AGE} & \textbf{ACCENTS} & \textbf{REGION} \\
    \midrule
p236 & 23 & English & Manchester \\
p248 & 23 & Indian &  \\
p261 & 26 & NorthernIrish & Belfast \\
p266 & 22 & Irish & Athlone \\
p280 & 25 & Unknown & France \\
p288 & 22 & Irish & Dublin \\
p301 & 23 & American & North Carolina \\
p312 & 19 & Canadian & Hamilton \\
p330 & 26 & American &  \\
p361 & 19 & American & New Jersey \\
    \bottomrule
\end{tabular}
\caption{\label{tab:female-speakers}Female Speakers.}
\end{table}

\end{document}